\documentclass[pra,twocolumn,aps,showpacs,superscriptaddress]{revtex4}
\usepackage{amsmath}
\usepackage{amssymb}
 \usepackage{latexsym}
\usepackage{float}
\usepackage{graphicx}
\newtheorem{theorem}{Theorem}

\begin{document}

\title{Quantum discord plays no distinguished role in characterization of complete 
positivity\,: Robustness of the traditional scheme} 
\author{Krishna Kumar Sabapathy}
\email{kkumar@imsc.res.in} 
\affiliation{Optics \& Quantum Information Group,\\ 
The Institute of Mathematical
  Sciences, C.I.T Campus, Tharamani, Chennai 600 113, India.}

\author{J. Solomon Ivan}
\affiliation{Indian Institute of Space Science and Technology, Valiamala, 
 Thiruvananthapuram 695 547, India.}

\author{Sibasish Ghosh}
\affiliation{Optics \& Quantum Information Group,\\ The Institute of Mathematical
  Sciences, C.I.T Campus, Tharamani, Chennai 600 113, India.}

\author{R. Simon}
\email{simon@imsc.res.in} 
\affiliation{Optics \& Quantum Information Group,\\ The Institute of Mathematical
  Sciences, C.I.T Campus, Tharamani, Chennai 600 113, India.}

\begin{abstract} 
{ The traditional scheme for realizing  open-system 
 quantum dynamics takes the initial state of the system-bath composite 
 as a simple product.
Currently, however, the issue of system-bath initial correlations possibly affecting the 
reduced dynamics of the system has been attracting considerable interest. 
The influential work of Shabani and Lidar\,[PRL {\bf 102}, 100402 (2009)] 
famously related this issue to quantum discord, a concept 
which has in recent years occupied the centre-stage of quantum information 
theory and has led to several fundamental results.    
 They  suggested that 
reduced dynamics is completely positive if and only if the initial 
system-bath correlations have vanishing quantum discord. Here we 
 show that there is, within the Shabani-Lidar framework,  
 no scope for any distinguished role for quantum discord in respect of 
complete positivity of reduced dynamics. 
Since most applications of quantum theory to real systems
rests on the traditional scheme, its robustness demonstrated here could be  
of far-reaching  significance.} 

\end{abstract}


\maketitle 
 
Every physical system is in interaction with its environment, {\em the bath}, to a smaller or 
larger degree of strength. As a 
consequence,  unitary Schr\"{o}dinger evolutions of the composite, the 
system plus the bath, manifests as dissipative 
non-unitary evolutions for the 
system of interest\,\cite{breuerbook}. The folklore scheme 
 for realizing such  open system 
dynamics is to elevate the  system states $\rho_S$ to the (tensor)  
products $\rho_S \otimes \rho_B^{\,\rm{fid}}$, for a {\em fixed} fiducial bath state 
$\rho_B^{\,\rm{fid}}$, then to evolve these uncorrelated system-bath states under a 
joint unitary $U_{SB}(t)$, and finally to trace out the bath degrees of 
freedom to obtain the evolved states $\rho_S(t)$ of the system\,:
\begin{align}
\rho_S \rightarrow \rho_S \otimes \rho_B^{\,\rm{fid}} 
&\to U_{SB}(t) \,\rho_S \otimes \rho_B^{\,\rm{fid}}\,U_{SB}(t)^{\dagger}\nonumber \\ 
\to \rho_S(t) &= \rm{Tr}_B\left[ U_{SB}(t) \,\rho_S 
 \otimes \rho_B^{\,\rm{fid}}\,U_{SB}(t)^{\dagger} \right].  
\label{e1}
\end{align} 
The resulting quantum dynamical process (QDP) $\rho_S \to 
\rho_S(t)$, parametrized by $\rho_B^{\,\rm{fid}}$ and $U_{SB}(t)$, is provably 
completely positive (CP)\,\cite{cp}. 

While every CP map can be thus realized with 
uncorrelated initial states of the composite, a suspicion that  more 
general realizations of CP 
maps  could be possible has 
always been lurking beneath the surface, and has occasionally erupted into 
passionate exchanges in the literature\,\cite{beyondcp}. 
 Possible effects of system-bath initial correlations on the  
 reduced dynamics for the system has been the subject of several recent 
 studies\,\cite{ini-corr,rosario08,shabani09}. 
  On the other hand, the concept of quantum discord\,\cite{zurek01,vedral01} (which tries to capture nonclassical correlations, even beyond entanglement) has come to   
 occupy  the centre-stage of quantum information 
theory for the past several years\,\cite{disc-imp1}, and  has led to many interesting results\,\cite{disc-imp2}.

A specific, carefully detailed, and precise formulation of the issue 
  of initial system-bath correlations possibly influencing the reduced dynamics was 
presented not long ago by Shabani and Lidar\,\cite{shabani09}  
(SL hereafter\,\cite{comment}). In this 
formulation, the distinguished bath state $\rho_B^{\,\rm{fid}}$ is replaced by 
a collection of (possibly correlated) system-bath 
 initial states $\Omega^{SB} \in {\cal B}({\cal H}_S 
\otimes {\cal H}_B)$, where ${\cal H}_S$, ${\cal H}_B$ are 
the Hilbert spaces of the system and bath, the  dimensions being $d_S$, $d_B$ respectively.
 The dynamics gets defined through a joint unitary $U_{SB}(t)$\,:
\begin{align} 
\rho_{SB}(0) \to \rho_{SB}(t) &= 
U_{SB}(t)\,\rho_{SB}(0)\,U_{SB}(t)^{\dagger}, \nonumber\\ 
 &~~~~~~~ \forall~ 
\rho_{SB}(0) \in \Omega^{SB}. 
\label{e2}
\end{align} 
This composite dynamics induces on the system the QDP 
\begin{align} 
\rho_S(0) \to \rho_S(t), \label{qdp}
\end{align} 
with $\rho_S(0)$ and $\rho_S(t)$ defined through this natural imaging from $\Omega^{SB}$ to the system state space $\Lambda_S$\,:  
\begin{align*} 
\rho_S(0) = \rm{Tr}_B\,\rho_{SB}(0),~\rho_S(t) = \rm{Tr}_B \rho_{SB}(t). 
\end{align*} 
It is evident that the folklore scheme obtains  as the special case 
$\Omega^{{SB}} = \{\,\rho_S \otimes \rho_B^{\,\rm{fid}}\,|\, 
\rho_B^{\,\rm{fid}}={\rm fixed}\,\}$.

This generalized formulation of QDP allows SL to transcribe the 
fundamental issue to this question: What are the necessary and sufficient 
conditions on the collection $\Omega^{SB}$ so that the induced QDP $\rho_S(0) 
\to \rho_S(t)$ in Eq.\,\eqref{qdp} is  
 guaranteed to be CP {\em for all} joint unitaries 
$U_{SB}(t)$? Motivated by the work of Rodriguez-Rosario et al.\,\cite{rosario08}, 
and indeed highlighting it 
 as {\em `a recent breakthrough'}, SL advance the following resolution to this 
 issue\,: 
\begin{theorem}{\rm (SL)}: The QDP in Eq.\,\eqref{qdp} is CP 
for all joint unitaries $U_{SB}(t)$ if and only if the  quantum 
discord vanishes for all $\rho_{SB} \in \Omega_{SB}$, i.e., if and only if the 
initial system-bath correlations are purely classical. 
\end{theorem}
The SL theorem has come to be counted among the more important recent results of quantum 
information theory, and has influenced an enormous number of authors. 

 In order that the QDP in Eq.\,\eqref{qdp} be {\em well defined} in the first place, the set $\Omega^{SB}$ 
 should necessarily satisfy the following two properties; since our entire analysis rests 
critically on these properties, we begin by motivating them.

{\em Property 1}: 
 No state $\rho_S(0)$ can have two (or more) pre-images in $\Omega^{SB}$.
To see this fact unfold, assume to the contrary that 
\begin{align*}
\rm{Tr}_B \rho_{SB}(0) &=\rm{Tr}_B \rho^{\,'}_{SB}(0),  
~~ \rho_{SB}(0) \neq \rho^{\,'}_{SB}(0),\nonumber\\
 & {\rm for~ two~ states~} \rho_{SB}(0), ~\rho^{\,'}_{SB}(0) \in \Omega^{SB}.
\end{align*}
It is clear that the difference $\triangle \rho_{SB}(0) = \rho_{SB}(0) - \rho^{\,'}_{SB}(0)$
should necessarily meet the property $\rm{Tr}_B \triangle \rho_{SB}(0)=0$. Let
$\{\lambda_u \}_{u=1}^{d_S^{\,2}-1}$ be a set of orthonormal hermitian
traceless $d_S \times d_S$ matrices so that together with the unit matrix
$\lambda_0 = 1\!\!1_{d_S \times d_S }$ these matrices form a basis for ${\cal
B}({\cal H}_S)$, the set of all $d_S \times d_S$  (complex) matrices.
Let $\{\gamma_v
\}_{v=1}^{d_B^{\,2}-1}$, $\gamma_0= 1\!\!1_{d_B \times d_B}$ be a similar
basis for ${\cal B}(\cal H_B)$. Then the $(d_Sd_B)^2$ tensor products
$\{\lambda_u \otimes \gamma_v \}$ form a basis for ${\cal B}({\cal H}_S
\otimes {\cal H}_B )$, and  $\triangle \rho_{SB}(0)$ can be written in the form
\begin{align*}
\triangle \rho_{SB}(0) = \sum_{u=0}^{d_S^{\,2}-1}\;
\sum_{v=0}^{d_B^{\,2}-1} C_{uv}\, \lambda_u \otimes \gamma_v, ~ C_{uv} \text{ real}.
\end{align*}
Now,
the property $\rm{Tr}_B\, \triangle \rho_{SB}(0)=0$ is strictly equivalent to
the demand that the expansion coefficient $C_{u0} = 0$, for all $u= 0,\,1,\,\cdots\, d_S^{\,2}-1$.
 Since the $[(d_S
d_B)^2-1]$-parameter unitary group $SU(d_Sd_B)$ acts {\em irreducibly} on the $[(d_S
d_B)^2-1]$-dimensional subspace of ${\cal B}({\cal H}_S \otimes {\cal H}_B)$
consisting of all traceless $d_Sd_B$-dimensional matrices [\,this is the adjoint
representation of $SU(d_Sd_B)$\,], there exists an $U_{SB}(t) \in SU(d_Sd_B)$
which takes $\triangle \rho_{SB}(0) \ne 0$ into a matrix whose
 expansion coefficient $C_{u0} \neq 0$ for
some $u$. That is, if the initial $\triangle \rho_{SB}(0)\ne 0$ then
one and the same system state $\rho_S(0)$ will evolve into two distinct
\begin{align*}
\rho_S(t) =& \rm{Tr}_B\left[ U_{SB}(t) \rho_{SB}(0) U_{SB}(t)^{\dagger}\right], 
     \nonumber\\
 \rho^{\,\prime} _S(t) =& 
\rm{Tr}_B\left[U_{SB}(t)\rho^{\,\prime} _{SB}(0)U_{SB}(t)^{\dagger}\right]
\end{align*} 
for some $U_{SB}(t)$, rendering the QDP
in Eq.\,\eqref{qdp} one-to-many, and hence ill-defined.

{\em Property 2}: 
  While every system state $\rho_S(0)$ need not have 
a pre-image {\em actually enumerated} in $\Omega^{SB}$, 
 the set of $\rho_S(0)$'s having pre-image should be sufficiently large.  
Indeed, Rodriguez-Rosario et al.\,\cite{rosario08} have
 rightly emphasised that it should be {\em `a large enough set of
states such that the QDP in Eq.\,\eqref{qdp} can be extended by linearity to all
states of the system'}. It is easy to see that 
if $\Omega^{SB}$ fails this property, 
 then the very issue of CP would make no sense. For, in carrying out verification
 of CP property, the QDP  would be required to act, 
as is well known\,\cite{choi75}, on 
 $\{|j\rangle \langle k|\}$ for $j,\,k = 1,2,\,\cdots\,d_S$; i.e., 
on generic complex $d_S$-dimensional
square matrices, and not just on positive or hermitian matrices alone.
Since the basic issue on hand is to check if the QDP as a map on
${\cal B}({\cal H}_S)$ is CP or not, it is essential that it be well
 defined (at least by linear extension)
on the entire {\em complex} linear space ${\cal B}({\cal H}_S)$.

With the two properties of $\Omega^{SB}$ thus motivated, we proceed to prove our main result.  We 
 `assume', {\em for the time being}, that every pure state $| \psi \rangle$ of 
the system has a pre-image in $\Omega^{SB}$. This assumption may appear, at first sight, 
to be a drastic one. But we show later that it entails indeed {\em no loss of generality}. 

 It is evident that, for every pure state $|\psi\rangle$, the pre-image 
 in $\Omega^{SB}$ has to necessarily 
 assume the (uncorrelated) product form $|\psi \rangle \langle \psi| \otimes 
\rho_B$ , $\rho_B$ being a state of the 
bath which could possibly 
depend on the system state $|\psi \rangle$.
Now, let $\{|\psi_k \rangle\}_{k=1}^{d_S}$ be an orthonormal basis in ${\cal H}_S$ 
and let $\{|\phi_{\alpha}\rangle\}_{\alpha=1}^{d_S}$ be another orthonormal 
basis related to the former through a complex Hadamard unitary matrix $U$. 
 Recall that a unitary U is Hadamard if $|U_{k\alpha}|= 1/\sqrt{d_S}$, independent 
of $k,\alpha$. For 
instance, the characters of the cyclic group of order $d_S$ 
 written out as a $d_S \times d_S$ matrix is 
Hadamard. The fact that the $\{|\psi_k\rangle \}$ basis and 
the $\{|\phi_{\alpha}\rangle \}$ basis are related by a Hadamard means that 
the magnitude of the inner product $\langle \psi_k |\phi_{\alpha} \rangle$ 
is independent of both $k$ and $\alpha$, and hence equals $1/\sqrt{d_S}$ uniformly.
 We may refer to such a pair as {\em relatively unbiased bases}. 
 
 
Let $|\psi_k \rangle \langle \psi_k| \otimes O_k$ be the pre-image of $|\psi_k 
\rangle \langle \psi_k|$ and $|\phi_{\alpha} \rangle \langle 
\phi_{\alpha}|\otimes \widetilde{O}_{\alpha}$ be that of $|\phi_{\alpha} \rangle 
\langle \phi_{\alpha}|$, $k,\alpha = 1,2,\cdots,d_S$. Possible dependence of 
the bath states $O_k$ on $|\psi_k \rangle$ and $\widetilde{O}_{\alpha}$ on 
$|\phi_{\alpha}\rangle$ has not been ruled out as yet.  
Since the maximally mixed system state can be expressed in 
two equivalent ways as $d_S^{-1} \sum_k |\psi_k \rangle \langle \psi_k| = 
d_S^{-1} \sum_{\alpha} |\phi_{\alpha} \rangle \langle \phi_{\alpha}|$, 
 {\em uniqueness} of its pre-image in $\Omega^{SB}$ demands (Property 1)
\begin{align*}
\sum_{k=1}^{d_S} |\psi_k \rangle \langle \psi_k|\otimes O_k 
= \sum_{\alpha=1}^{d_S} |\phi_{\alpha} \rangle \langle \phi_{\alpha}| 
  \otimes \widetilde{O}_{\alpha}. 
\end{align*} 
Taking projection of both sides on 
$|\psi_j \rangle \langle \psi_j|$, and using $|\langle \psi_j|\phi_{\alpha} 
\rangle|^2 = d_S^{-1}$, we have 
\begin{align*} 
O_j = \frac{1}{d_S} 
\sum_{\alpha=1}^{d_S} \widetilde{O}_{\alpha}, ~~ j=1,2,\cdots,d_S, 
\end{align*} 
while projection on $|\phi_{\beta} \rangle \langle \phi_{\beta}|$ leads to 
\begin{align*} 
\widetilde{O}_{\beta} = \frac{1}{d_S} \sum_{k=1}^{d_S} O_k, ~~ 
\beta=1,2,\cdots,d_S. 
\end{align*} 
These $2 d_S$ constraints together imply that $O_j= \widetilde{O}_{\beta}$ 
uniformly for all $j,\,\beta$.  
Thus the pre-image of $ |\psi_k \rangle 
\langle \psi_k|$ is $ |\psi_k \rangle \langle \psi_k| \otimes 
\rho_B^{\,\rm{fid}}$ and that of $|\phi_{\alpha} \rangle \langle 
\phi_{\alpha}|$ is $ |\phi_{\alpha} \rangle \langle \phi_{\alpha}| \otimes 
\rho_B^{\,\rm{fid}}$, for all $k,\alpha$, for some {\em fixed 
 bath state} $\rho_B^{\,\rm{fid}}$.  
And, perhaps more importantly, the 
pre-image of the maximally mixed state $d_S^{-1} 1\!\!1$ 
 necessarily equals the product $d_S^{-1} 1\!\!1 \otimes \rho_B^{\,\rm{fid}}$.

Taking another pair of relatively unbiased bases $\{|\psi^{\,'}_k \rangle\}$, 
$\{ |\phi^{\,'}_{\alpha}\rangle\}$ one similarly concludes that the pure 
states $ |\psi^{\,'}_k \rangle \langle \psi^{\,'}_k|$, $ |\phi^{\,'}_{\alpha} 
\rangle \langle \phi^{\,'}_{\alpha}|$ too have pre-images $|\psi^{\,'}_k 
\rangle \langle \psi^{\,'}_k| \otimes \rho_B^{\,\rm{fid}}$, 
$|\phi^{\,'}_{\alpha} \rangle \langle \phi^{\,'}_{\alpha}| \otimes 
\rho_B^{\,\rm{fid}}$ respectively, with the same 
fixed fiducial bath state $\rho_B^{\,\rm{fid}}$. This is so, since the 
maximally mixed state is {\em common} to both sets.   

Considering in this manner enough  
 number of pure states or projections $|\psi \rangle \langle \psi|$ sufficient to span---by linearity---the entire system 
state space $\Lambda_S$, and hence ${\cal B}({\cal H}_S)$,
 one readily concludes that {\em every element} of $\Omega^{SB}$ 
 {\em necessarily} needs to be 
of the product form $\rho_S(0) \otimes \rho_B^{\,\rm{fid}}$, 
for some {\em fixed}  
bath state $\rho_B^{\,\rm{fid}}$. But this is exactly the folklore 
realization of non-unitary dissipative dynamics given in Eq.\,\eqref{e1}, to surpass 
which was the primary goal of the SL scheme. We have thus proved our principal result\,: 
\begin{theorem}:
No initial correlations---even classical ones---are 
permissible within the SL scheme.
\end{theorem}

As we have noted, if at all a pure state $\rho_S(0)= |\psi \rangle \langle \psi|$ has a pre-image in 
$\Omega^{SB}$ it would necessarily be of the product form $ |\psi \rangle 
\langle \psi| \otimes \rho_B$, for some (possibly $|\psi\rangle$-dependent) 
bath state $\rho_B$. While this is self-evident and is independent of SL, 
 it is instructive to view it  as a consequence of 
the {\em necessary condition part} of SL theorem. Then our 
principal conclusion above can be rephrased to say that validity of SL theorem for 
pure states of the system readily leads to the folklore product-scheme as the {\em only solution} within the SL framework.   
This interesting aspect comes through  in an even  
more striking manner in our proof below   
  that our earlier `assumption' is one without loss of generality.

{\em Our assumption entails no loss of generality}:  
Let us focus, to begin with, on the convex hull  $\overline{\Omega^{SB}}$ of
$\Omega^{SB}$ rather than the full (complex) linear span of $\Omega^{SB}$   
to which we are entitled. Let us further assume that the image 
of $\overline{\Omega^{SB}}$ under the convexity-preserving 
linear imaging (projection) map $\rho_{SB}(0) \to 
{\rm Tr}_B \rho_{SB}(0)$ fills not the entire (convex) state 
space---the $(d^{\,2}_S-1)$-dimensional 
generalized Bloch sphere $\Lambda_S$---of the system, but 
only a portion thereof, possibly a very small part. Even so, in order that 
our QDP in Eq.\,\eqref{qdp} be well-defined, this portion would 
{\em occupy a non-zero volume} of the $(d^{\,2}_S-1)$-dimensional 
 state space $\Lambda_S$ of the system (Property 2).

\begin{figure}
\scalebox{0.8}{
\includegraphics{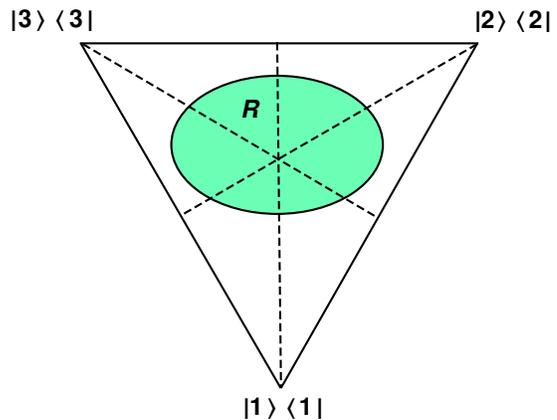}}
\caption{Depicting, for the case $d_S=3$ (qutrit), the image of $\overline{\Omega^{SB}}$ under 
     ${\rm Tr}_B (\cdot)$ in the plane spanned by the commuting (diagonal)  
      $\lambda$-matrices $(\lambda_3,\lambda_8)$.\\
 \label{fig2}} 
\end{figure}

Let us consider one set of all mutually commuting elements of $\Lambda_S$. If 
the full state space $\Lambda_S$ were available under the imaging 
$\rho_{SB}(0) \to 
{\rm Tr}_B \rho_{SB}(0)$ of $\overline{\Omega^{SB}}$, 
then the resulting mutually commuting images would have  
filled the  entire $(d_S-1)$-simplex, 
the classical state space of a $d_S$-level system, 
this being respectively the triangle and the tetrahedron 
when $d_S=3,4$\,\cite{simplex}. 
Since the full state space is assumed to be not available as image of $\overline{\Omega^{SB}}$, these commuting 
elements fill a, possibly very small but of nontrivial measure, 
proper convex subset of the $(d_S-1)$-simplex, 
 depicted in Fig.\,\ref{fig2} as region $R$ for the case $d_S=3$ (qutrit).

Elements of these simultaneously diagonal density matrices 
of the system can be expressed as convex 
sums of pure states or one-dimensional 
projections. 
For a generic element in this region, the spectrum is 
non-degenerate, and hence the projections are unique and commuting, being the 
eigenstates of $\rho_S(0)$, and correspond to the $d_S$ vertices of the ($d_S-1$)-simplex. In the case of qutrit, it is pictorially seen in Fig.\,\ref{fig2} that only the points  
 on the bisectors (the three dotted lines) correspond to doubly degenerate density 
matrices and the centre alone is 
triply degenerate, rendering transparent the fact that 
being nondegenerate is a generic attribute of  region $R$.  

Now  consider the pre-image $\rho_{SB}(0)$ in $\overline{\Omega^{SB}}$ of 
such a 
non-degenerate $\rho_S(0) \in R$. Application of the SL requirement 
of vanishing discord (again, only the necessity part of 
the SL theorem) to this $\rho_{SB}(0)$ implies
that  this pre-image  has  the form\,\cite{zurek01,modirmp}
\begin{align}
\rho_{SB}(0) = \sum_{j=1}^{d_S} p_j |j \rangle \langle j| \otimes \rho_{Bj}(0), 
\label{e4}
\end{align}
 where the probabilities $p_j$ and the pure states $|j \rangle \langle j|$
  are uniquely determined (in view of nondegeneracy) by the spectral resolution  
\begin{align*} 
\rho_S(0) = {\rm 
Tr}_B\,\rho_{SB}(0) = \sum_{j=1}^{d_S} p_j |j \rangle \langle j| . 
\end{align*} 
And $\rho_{Bj}(0)$'s are bath states, possibly dependent on $|j \rangle 
\langle j|$  as indicated by the label $j$ in $\rho_{Bj}(0)$.  
  These considerations  hold for {\em every}  nondegenerate element of  
  region $R$ of probabilities $\{\,p_j\,\}$. 
   In view of generic nondegeneracy,  
 the requirement\,\eqref{e4}  
 implies that each of the $d_S$ pure states $|j \rangle \langle j|$ has 
 pre-image of the form $|j \rangle \langle j|\otimes \rho_{Bj}(0)$ in the {\em linear span} of the pre-image of R---at least   
 as seen by the evolution\,\eqref{e2}. That is 
$\rho_{Bj }(0)$'s cannot be dependent on the probabilities $\{\,p_j\}$. 

Since every pure state of the system 
constitutes one of the vertices of some  
$(d_S-1)$-simplex comprising one set of all mutually commuting density 
operators $\rho_S(0)$, the conclusion that a pure 
state {\em effectively} has  in the linear span of $\Omega^{SB}$ a pre-image, and one necessarily of the product form, 
{\em applies to every pure state}, showing that the `assumption' 
in our earlier analysis indeed {\em entails no loss of generality}.

To summarize, it is clear that the dynamics described 
 by\,\eqref{e2} and\,\eqref{qdp} would `see' only the  full (complex) linear span of  
 $\Omega^{SB}$, and {\em not so much the actual enumeration}
 of $\Omega^{SB}$  as such. This is notwithstanding the fact that, 
 as indicated by the projection map $\rho_{SB}(0) \to \rho_S(0) = \rm{Tr}_B\,\rho_{SB}(0)$,  the 
only elements of this linear span 
which are immediately relevant for the QDP are those which are 
hermitian, positive semidefinite, and have unit trace. 
 Since no system state can have two or more pre-images (Property 1), 
 in order that the QDP in\,\eqref{qdp}
be well defined these relevant elements are forced to 
constitute a {\em faithful linear embedding}  
of (a nontrivial convex subset of) the system's state space $\Lambda_S$ in  ${\cal B}({\cal H}_S \otimes {\cal H}_B)$. 
 In the SL scheme of things, this leaves us with just the folklore embedding
 $\rho_S(0)\to \rho_{SB}(0)=\rho_S(0)\otimes \rho_B^{\,{\rm fid}}$. 
 This is the principal conclusion that emerges.

Let us view this from a slightly different position.
Since there is no conceivable 
manner in which a linear map acting on elements of $\Omega^{SB}$ 
 could be prevented from acting on 
convex sums (indeed, the linear span) of such elements, we may assume---without loss of 
of generality---$\Omega^{SB}$ to be convex and ask,  
consistent with  the SL theorem: 
  What are the possible choices for  
 the set $\Omega^{SB}$ to be {\em convex and at the same time consist entirely of states 
of vanishing quantum discord}. 
 One possibility comprises elements of the form 
$\rho_{SB}(0) = \rho_{S}(0)\otimes \rho_B^{\,{\rm fid}}$, for a fixed bath state 
 $\rho_B^{\,{\rm fid}}$  and arbitrary system state $\rho_{S}(0)$. 
This is recognized to be simply the folklore  case. 
The second one consists of elements of the form 
$\rho_{SB}(0) = \sum_{j} p_j |j \rangle \langle j| \otimes \rho_{Bj}(0)$,  
 for a {\em fixed (complete) set of orthonormal pure states} 
$\{|j \rangle \langle j|\}$, a  
 case  restricted to {\em  mutually commuting density operators} of the system.   
  This seems to be the case studied by Rodriguez-Rosario et al.\,\cite{rosario08},
  but the very notion of CP is unlikely to make much sense in this non-quantum case of  
 {\em classical state space} (of dimension $d_S-1$ rather than 
$d_S^{\,2}-1$), the honorific `a recent breakthrough' notwithstanding. 

The stated goal of SL  was to give 
a {\em complete characterization}
 of possible initial correlations that lead to CP maps. 
It is possibly in view of the belief that there was 
a large class of permissible initial correlations out there 
within the SL framework, and that that class now stands  
fully characterized by the SL theorem\,\cite{shabani09}, that
a large number of recent papers tend to list complete characterization of   
CP maps among the principal achievements of quantum discord\,\cite{disc-cp}. 
Our result implies, with no irreverence whatsoever to quantum discord, 
 that characterization of CP maps may not yet be 
 rightfully paraded as one of the achievements of quantum discord.


The SL theorem has influenced an enormous number 
of authors, and it is possible that those results of these 
authors which make essential use of the sufficiency part of the SL theorem need recalibration in the light of our result.

There are other, potentially much deeper, implications of our finding. 
 Our analysis---strictly  within the SL  
framework---has shown that this framework brings one exactly back to the 
  folklore scheme itself, as if it were a {\em fixed point}.  
 This is not at all a negative result for two reasons. First,  
 it shows that quantum discord is no `cheaper' (to accommodate) than entanglement as far as 
 complete positivity of QDP is concerned. Second, and more importantly, the
fact that the folklore product-scheme survives  attack 
 under this well-defined and fairly general SL framework demonstrates  
 its, perhaps unsuspected, {\em robustness}. In view of the fact that 
 this scheme has been at the heart of most applications 
 of quantum theory to real situations, virtually in every area of physical 
science, and even beyond, its robustness the SL framework 
helps to demonstrate is likely to prove to be of far-reaching significance.




\end{document}